\documentstyle[12pt,thmsa,sw20aip]{article}

\input{tcilatex}
\input psfig
\begin{document}

\author{M.B.Altaie\thanks{%
E-mail: maltaie@yu.edu.jo} \and Department of Physics, Yarmouk University,
Irbid-Jordan}
\title{Back-reaction of quantum fields in an Einstein Universe.}
\date{April 2001}
\maketitle

\begin{abstract}
We study the back-reaction effects of the finite-temperature massless scalar
field and the photon field in the background of the static Einstein
universe. In each case we find a relation between the temperature of the
universe and its radius.These relations exhibit a minimum radius below which
no self-consistent solution for the Einstein field equation can be found. A
maximum temperature marks the transition from the vacuum dominated era to
the radiation dominated era. An interpretation to this behavior in terms of
Bose-Einstein condensation in the case of the scalar field is given.
\end{abstract}

\section{Introduction}

Many authors have investigated the behavior of quantum fields in curved
spacetimes (for a thorough in-depth review see ref. [1]). These
investigations came in an endeavor to understand the origin of the universe
and the creation of matter, presumably, out of an arbitrary state of nothing
(the vacuum). The subject was initiated by the discovery of Penzias and
Wilson [2]of the microwave background radiation, where it was observed that
the galaxies swim in a global cold bath at about $2.73$K. The source of this
radiation was found to be cosmic; therefore, it was called the Cosmic
Microwave Background radiation (CMB). This radiation was found to be
isotropic over large angular scale of observation, and that it has a Planck
spectrum for a radiating black body at about $2.73$K.

The discovery of the CMB revived the theory of the hot origin of the
universe (the big-bang model) which was worked out in the late 1940's by
Gamow and his collaborators. The most refined analysis along this line
predicted a cosmic background radiation at a temperature about $5$K (for a
concise recent review of the subject see ref. [3]) . Therefore the Penzias
and Wilson discovery was considered a good verification of what was called
the big bang model. However, since the Gamow model started with the universe
at the times when the temperature was about $10^{12}$K, the new interest in
the origin of the universe sought much earlier times at much higher
temperatures. The new interest arose in studying the state of the universe
in the period from near the Planck time ($\sim 10^{-44}s$) to the grand
unification time ($\sim 10^{-34}s$). This is the era when quantum effects
played a decisive role in the subsequent developments of the universe, and
it is also the era when particle processes could have left permanent
imprints on the content of the universe.

The works dealing with this question started by mid 1970's when matter
fields where brought into connection with spacetime curvature through the
calculation of the vacuum expectation value of the energy-momentum tensor $%
<0|T_{\mu \nu }|0>$ \ [4-8].\thinspace The motivations for studying this
quantity stems from the fact that $T_{\mu \nu }$ is a local quantity that
can be defined at a specific spacetime point, contrary to the particle
concept which is global. The energy-momentum tensor also acts as a source of
gravity in the Einstein field equations, therefore $<0|T_{\mu \nu }|0>$
plays an important role in any attempt to model a self-consistent dynamics
involving the classical gravitational field coupled to the quantized matter
fields. So, once $<0|T_{\mu \nu }|0>$ is calculated in a specified
background geometry, we can substitute it on the RHS of the Einstein field
equation and demand self consistency, i.e 
\begin{equation}
R_{\mu \nu }-\frac{1}{2}g_{\mu \nu }R=-8\pi <0|T_{\mu \nu }|0>,  \label{q1}
\end{equation}
where $R_{\mu \nu }$ is the Ricci tensor, $g_{\mu \nu }$ is the metric
tensor and $R$ is the scalar curvature.

The solution of (\ref{q1}) will determine the development of the spacetime
in presence of the given matter field, for which $|0>$ can be unambiguously
defined. This is known as the ''back-reaction problem''. It is interesting
to perform the calculation of $<0|T_{\mu \nu }|0>$ in
Friedman-Robertson-Walker (FRW) models because the real universe is, more or
less, a sophisticated form of the Friedman models. However the
time-dependence of the spacetime metric generally creates unsolvable
fundamental problems. One such a problem was the definition of vacuum in a
time-dependent background [9]; a time-dependent background is eligible for
producing particles continuously, therefore, pure vacuum states in the
Minkowskian sence do not exist. Also an investigation into the
thermodynamics of a time-dependent systems lacks the proper definition of
thermal equilibrium, which is a basic necessity for studying
finite-temperature field theory in curved backgrounds.[10].

Of all the available solutions of the Einstein field equations, the static
Einstein universe stands the above two funamental challenges. First, being
static, the Einstein universe leaves no ambiguity in defining the vacuum
both locally and globally [1]. The same feature also allows for thermal
equilibrium to be defined unambiguously. Furthermore, the Einsein static
metric is conformal to all Robertson-Walker metrics, and it was shown by
Kennedy [10] that thermal Green functions for the static Einstein universe
and the time-dependent Robertson-Walker universe are coformally related,
hence deducing a (1-1) correspondance between the vacuum and the many
particle states of both universes. So that, under the equilibrium condition,
the thermodynamics of quantum fields in an Einstein universe of radius $a$
is equivalent to that of an instantaneously static Friedman-Roberson-Walker
(FRW) universe of equal radius [4,7,11]. This means that the results
obtained in FRW universe would be qualitatively the same as those obtained
in an Einstein universe.

Dowker and Critchley [12] considered the finite temperature corrections for
the massless scalar field in an Einstein universe using the technique of
finite-temperature Green's functions. Later Altaie and Dowker [13]
calculated the finite temperature corrections to the massless scalar field,
the neutrino field and, the photon field in the background of an Einstein
universe. The results of the calculation for the photon field was used to
deduce a self-consistent solution for the Einstein field equation, i.e, a
back-reaction problem, from which a relation between the temperature and the
radius of the Einstein universe was deduced. However, this relation was not
fully exploited at that time and therefore some of the thermodynamical
aspects were kept unexposed. Hu [11] considered the effects of
finite-temperature comformally coupled massless scalar field in a closed
Robertson-Walker universe using the results of Altaie and Dowker [13], and
assuming that the thermal equilibrium is established for the scalar
particles throughout the history of the universe. In the high-temperature
limit Hu found that the universe expands linearly in cosmic time near the
singularity. In the low-temperature limit, it reduces to the Starobinsky-de
Sitter type solution where the singularity is avoided in an exponential
expansion, concluding that the finite-temperature formalism provides a
unifying framework for the description of the interplay of vacuum and
radiation energy and their combined effect on the state of the early
universe.

Recently Plunien et al [14] considered the dynamical Casimir effect at
finite temperature. They reported that finite temperatures can enhance the
pure vacuum effect by several orders of magnitude. Although the relevance of
this result was addressed in the context of an effort aiming at the
experimental verification of the Casimir effect, it does have a useful
implication in respect of the theoretical understanding of the finite
temperature corrections to the vacuum energy density in closed spacetimes.

In this letter we will reconsider the calculation of the back-reaction
effect of the conformally coupled massless scalar field and the photon field
in the background of the Einstein static universe. The aim is to expose the
thermal behavior of the system, analyze and interpret details that may have
been overlooked in previous studies, and investigate the possibility of
assigning any practical applicability of the results.

\section{The vacuum energy density and Back-reaction}

{\it \ }The metric of the Einstein static universe is given by 
\begin{equation}
ds^{2}=dt^{2}-a^{2}\left[ d\chi ^{2}+\sin ^{2}\chi \left( d\theta ^{2}+\sin
^{2}\theta d\phi ^{2}\right) \right] \,,  \label{q2}
\end{equation}
where $a$ is the radius of the spatial part of the universe $S^{3}$ and, $%
0\leq \chi \leq \pi $, $0\leq \theta \leq \pi $, and $0\leq \phi \leq 2\pi $.

We consider an Einstein static universe being filled with a massless boson
gas in thermal equilibrium at temperature $T$. The total energy density of
the system can be written as

\begin{equation}
<T_{00}>_{tot}=<T_{00}>_{T}+<T_{00}>_{0},  \label{q3}
\end{equation}
where $<T_{00}>_{0}$ is the zero-temperature vacuum energy density and $%
<T_{00}>_{T}\,$is the corrections for finite temperatures, i.e.,

\begin{equation}
<T_{00}>_{T}=\frac{1}{V}\sum_{n}\frac{d_{n}\epsilon _{n}}{\exp \beta
\epsilon _{n}-1},  \label{q4}
\end{equation}
where $\epsilon _{n}$ and $d_{n}$ are the eigen energies and degeneracies of
the $n$th state, and $V$=$2\pi ^{2}a^{3}$ is the volume of the spatial
section of the Einstein universe.

To investigate the back-reaction effect of finite-temperature quantum fields
on the behavior of the spacetime we should substitute for $<T_{00}>_{tot}$
on the RHS of the Einstein field, but this time with the cosmological
constant $\lambda $, i.e.

\begin{equation}
R_{\mu \nu }-\frac{1}{2}g_{\mu \nu }R+g_{\mu \nu }\lambda =-8\pi <T_{\mu \nu
}>_{tot}.  \label{q5}
\end{equation}

In order to eliminate $\lambda $ from (\ref{q5}) we multiply both sides with 
$g_{\mu \nu }$ and sum over $\mu $ and $\nu $, then using the fact that $%
T_{\mu }^{\mu }=0$ for massless fields, and for the Einstein universe $%
R_{00}=0$ , $g_{00}=1$ ,and $R=\frac{6}{a^{2}}$, we get

\begin{equation}
\frac{6}{a^{2}}=32\pi <T_{00}>_{tot}.  \label{q6}
\end{equation}

\subsection{Scalar Field.}

For a conformally coupled massless scalar field the zero-temperature vacuum
energy density in an Einstein universe is [4,6]

\begin{equation}
<T_{00}>_{0}=\frac{1}{480\pi ^{2}a^{4}}  \label{q7}
\end{equation}

The eigen-energies and degeneracies are $\epsilon _{n}=\frac{n}{a}$ and $%
d_{n}=n^{2}$ resectively, so that (\ref{q3}) gives

\begin{equation}
<T_{00}>_{tot}=\frac{1}{2\pi ^{2}a^{4}}\sum_{n=1}^{\infty }\frac{n^{3}}{\exp
(n/Ta)-1}+\frac{1}{480\pi ^{2}a^{4}}.  \label{q8}
\end{equation}

Using this mode-sum expression, Altaie and Dowker [13] calculated the finite
temperature corrections for the vacuum energy density of the conformally
coupled massless scalar field in the Einstein universe. The results which
are functions of a single parameter $\xi (=Ta)$, were then subjected to the
high and low-temperature limits. It was found that in the low-temperature
(or small radius) limit the zero-temperature vacuum energy density is
recovered, i.e

\begin{equation}
\stackunder{\xi \longrightarrow 0}{\lim }<T_{00}>_{tot}=\frac{1}{480\pi
^{2}a^{4}},  \label{q9}
\end{equation}
and in the high-temperature (or large radius) limit the behavior of the
system is totally Planckian,

\begin{equation}
\stackunder{\xi \longrightarrow \infty }{\lim }<T_{00}>=\frac{\pi ^{2}}{30}%
T^{4}.  \label{q10}
\end{equation}

In order to investigate the back-reaction effect of the field we substitute
for $<T_{00}>_{tot}$ from (\ref{q8}) in (\ref{q6}) and request a
self-consistent solution, we get

\begin{equation}
a^{2}=\frac{8}{3\pi }\sum_{n=1}^{\infty }\frac{n^{3}}{\exp (n/Ta)-1}+\frac{1%
}{90\pi }.  \label{q11}
\end{equation}

This equation determines a relation between the temperature $T$ and the
radius $a$ of the Einstein universe in presence of the conformally coupled
massless scalar field. The solutions of this equation are shown in Fig.
1.Two regimes are recognized: one corresponding to small values of $\xi $
where the temperature rises sharply reaching a maximum at $T_{max}\approx
2.218T_{p}=3.15\times 10^{32}$K at a radius $a_{t}\approx
0.072l_{p}=1.16\times 10^{-34}cm$. Since this regime is controlled by the
vacuum energy (the Casimir energy), therefore we prefer to call it the
''Casimir regime''. The second regime is what we call the ''Planck regime'',
which correspond to large values of $\xi $, and in which the temperature
asymptotically approaches zero for very large values of $a$. This behavior
was over looked by Hu [11].

\begin{figure}[ht]
\centerline{\psfig{file=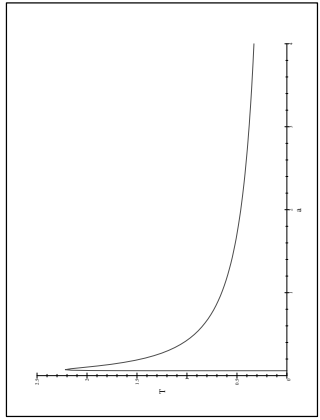,width=1.8524in,height=2.3903in}}
\caption{The temperature-radius relationship for massless scalar field in an Einstein universe}
\end{figure}

From (\ref{q11}) it is clear that at $T=0$ the radius of an Einstein
universe has a minimum value $a_{0}$, below which no consistent solution of
the Einstein field equation exist. This is given by

\begin{equation}
a_{0}=\left( \frac{1}{90\pi }\right) ^{1/2}l_{p}.  \label{q12}
\end{equation}

From (\ref{q6}) and (\ref{q10}) we can calculate the background (Tolman)
temperature of the universe in the limit of large radius. This is given by

\begin{equation}
T_{b}=\left( \frac{45}{8\pi ^{3}a^{2}}\right) ^{1/4},  \label{q13}
\end{equation}
for example, at $a=1.38\times 10^{28}cm$ we obtain $T=31.556$K.

Conversily if we demand that the background temperature be the presently
measured value of $2.73$K, the radius of the universe should be $%
1.\,294\times 10^{30}cm$. This is about two orders of magnitude larger than
the estimated Hubble length of $1.38\times 10^{28}cm$.

\subsection{Photon Field}

The vacuum energy density of this field at zero temperature is given by [6]

\begin{equation}
<T_{00}>_{0}=\frac{11}{240\pi ^{2}a^{4}}.  \label{q14}
\end{equation}

The total energy density of the system in terms of the mode-sum can be
written as

\begin{equation}
<T_{00}>_{tot}=\frac{1}{\pi ^{2}a^{4}}\sum_{n=2}^{\infty }\frac{n(n^{2}-1)}{%
\exp (n/Ta)-1}+\frac{11}{240\pi ^{2}a^{4}}  \label{q15}
\end{equation}

In the low-temperature limit the result reduce to [13]

\begin{equation}
\lim_{\xi \longrightarrow 0}<T_{00}>_{tot}=\frac{11}{240\pi ^{2}a^{4}}.
\label{q16}
\end{equation}

Substituting this in (\ref{q6}) we get

\begin{equation}
a_{0\gamma }=\left( \frac{11}{45\pi }\right) ^{1/2}l_{p}.  \label{q17}
\end{equation}

This is the minimum radius for an Einstein static universe filled with
photons at finite temperatures.

In the high temperature (or large radius) limit the result is

\begin{equation}
\lim_{\xi \longrightarrow \infty }<T_{00}>_{tot}=\frac{\pi ^{2}}{15}T^{4}-%
\frac{1}{6}\frac{T^{2}}{a^{2}}.  \label{q18}
\end{equation}

The back-reaction of the field can be studied if we substitue (\ref{q15}) in
(\ref{q6}) where this time we obtain

\begin{equation}
a^{2}=\frac{16}{3\pi }\sum_{n=2}^{\infty }\frac{n(n^{2}-1)}{\exp (n/Ta)-1}+%
\frac{11}{45\pi }.  \label{q19}
\end{equation}

The solutions of this equation are dipicted in Fig. 2, where we see that the
behavior is qualitatively the same as that encountered in the conformally
coupled scalar field case. The minimum radius permissable for a
self-consistent solution to exist in presence of the photon field is $%
a_{0}=0.279l_{p},$ and the maximum temperature $T_{\max
}=1.015T_{p}=1.44\times 10^{32}$ K at $a_{t}=0.34l_{p}=5.5\times 10^{-34}$ $%
cm$.

The background (Tolman) temperature of the photon field is

\begin{equation}
T_{b\gamma }=\left( \frac{45}{16\pi ^{3}a^{2}}\right) ^{1/4}.  \label{20}
\end{equation}

At the radius of $1.38\times 10^{28}cm$ we obtain a background temperature
of $30.\,267$ K, and if we require that the background temperature to be the
same as the avarage measured value of $2.73$ K, the radius of the Eistein
universe has to be $1.83\times 10^{30}cm$. Again more than two orders of
magnitude larger than the estimated value of Hubble length.

\begin{figure}[ht]
\centerline{\psfig{file=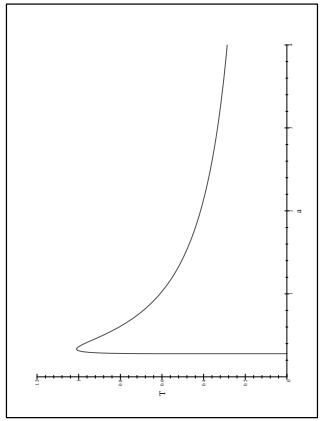,width=1.8524in,height=2.3903in}}
\caption{The temperature-radius relationship for the photon field in an Einstein universe.}
\end{figure}

\section{Bose-Einstein Condensation}

{\it \ }In an earlier work [15], we studied the BEC of non-relativistic spin
0 and spin 1 particles in an Einstein universe. We found that the finiteness
of the system resulted in smoothing-out the singularities of the
thermodynamic functions which are normally found in infinite systems, so
that the phase transitions in curved space becomes non-critical. We also
remarked the enhancement of the condensate fraction and the displacement of
the specific-heat maximum toward higher temperatures. Singh and Pathria [16]
considered the BEC of a relativistic conformally coupled massive scalar
field. Their results confirmed our earlier findings of the non-relativistic
case. Recently [17] we considered the BEC of the relativistic massive spin-1
field in an Einstein universe. Again the resuls confirmed our earlier
findings concerning the general features of the BEC in closed spacetimes. So
the above mentioned features became an established general features of the
BEC of quantum fields in curved spaces.

Parker and Zhang [18], considered the ultra-relativistic BEC of the
minimally coupled massive scalar field in an Einstein universe in the limit
of high temperatures. They showed, among other things, that
ultra-relativistic BEC can occur at very high temperature and densities in
the Einstein universe, and by implication in the early stages of a
dynamically changing universe. Parker and Zhang [19], also showed that the
Bose-Einstein condensate can act as a source for inflation leading to a
de-Sitter type universe. However, Parker and Zhang gave no specific value
for the condensation temperature of the system.

Here we are going to use the ready result obtained for the condensation
temperature $T_{c}$ of the conformally coupled massless scalar field in
order to explain the change of behavior of the system from the Casimir
(vacuum dominated) regime to the Planck (radiation dominated) regime. This
change wich is taking place at a well defined maximum temperature which we
called the transition temperature $T_{t}$.

The condensation temperature of the conformally coupled massive scalar field
is given by[16,20]

\begin{equation}
T_{c}=\left( \frac{3q}{m}\right) ^{1/2}\left( 1+\frac{1}{m^{2}a^{2}}\right)
^{-1/4}.  \label{q21}
\end{equation}

In the massless limit this gives

\begin{equation}
T_{c}=\sqrt{3qa},  \label{q22}
\end{equation}
where $q$ is the number density of the particles.

For the Einstein universe specifically we have shown that (see Eq. \ref{q6})

\begin{equation}
<T_{00}>_{tot}=\frac{3}{16\pi a^{2}}.  \label{q23}
\end{equation}

This means that, if the back-reaction of the field is to be taken into
consideration, the number density of the particles in the system $q$ will be
inversely proportional to $a^{2}$, this enable us to write

\begin{equation}
qa^{2}=\text{constant,}  \label{q24}
\end{equation}
for any values of $q$ and $a$. This means that 
\begin{equation}
q_{p}=q_{0}\left( \frac{a_{0}}{a_{p}}\right) ^{2}.  \label{q25}
\end{equation}

Substituting this in (\ref{q22}) we get the expression for the condensation
temperature of the conformally coupled massless scalar field in Einstein
universe as

\begin{equation}
T_{c}=\sqrt{\frac{3q_{p}a_{p}^{2}}{a_{0}}.}  \label{q26}
\end{equation}

The estimated upper bound on the net average particle number density of the
universe at present is $q_{p}<10^{-24}cm^{-3}$ (see Dolgov and Zeldovich
[21]). If this upper bound is adopted, then using (\ref{q26}) we can
calculate the condensation temperature of the conformally coupled field at
any specified radius. If we substitue for the radius of the Einstein
universe the estimated value of Hubble length, i.e $a_{p}=1.8\times
10^{28}cm $, then we can write

\begin{equation}
T_{c}=\frac{1.268}{\sqrt{a_{0}}}.  \label{q27}
\end{equation}

We have already found that the transition from the Casimir (vacuum) regime
to the Planck (radiation) regime takes place at $a_{0}=$ $0.072l_{p}$.
Substituting this in (\ref{q27}) we get

\begin{equation}
T_{c}=\allowbreak 4.\,725T_{p}.  \label{q28}
\end{equation}

Clearly we obtain a condensation temperature which is the same order of
magnitude as the transition temperature obtained earlier for the conformally
coupled massless scalar particles.

This strongly suggests that the transition from the Casimir (vacuum) regime
into the Planck regime is taking place as a result of Bose-Einstein
condensation of the vacuum energy so that as the condensate is formed a free
absorbsion and emission of massless quanta by the condensate is expected to
take place and the system will start behaving according to Planck law.

\section{Conclusions}

In conclusion we can say that the present study exhibited some features of
the thermodynamical behavior of the Einstein universe. The main finding are

1. The thermal development of the universe is a direct consequence of the
state of its global curvature.

2. The universe avoids the singularity at $T=0$ through quantum effects (the
Casimir effect) because of the non-zero value of $<T_{00}>_{0}$. A non-zero
expectation value of the vacuum energy density always implies a symmetry
breaking event.

3. During the Casimir regime the universe is totally controlled by vacuum
and this region represent the vacuum era. The energy content of the universe
is a function of its radius. Using the conformal relation between the static
Einstein universe and the closed FRW universe [10], this result indicates
that in a FRW\ model there would be a continuous creation of energy out of
vacuum as long as the universe is expanding , a result which is confirmed by
Parker long ago [22]. The steep, nearly vertical line in Fig. 1 and 2.
suggests that the universe started violently and had to relax later.

4. At high temperatures a new quantum-thermal effects do interfere causing a
phase transition at about $T_{max}=2.218T_{p}=3.15\times 10^{32}$K for the
massless scalar field, and at $T_{max}=1.015T_{p}=1.44\times 10^{32}$ K for
the photon field. The calculations show that a Bose-Einstein condensation of
massless quanta (at least in the scalar field case) may be responsible for
the transition. The values of these peaks agrees with the expectations of
particle physics in respect to the the era of total unification of forces.

The recent findings of Plunien etal [14] that finite temperatures can
enhance the pure vacuum effect by several orders of magnitude, can be used
to explain the behavior of our system during the Casimir (vacuum) regime.
Since this means that the finite temperature corrections will surely enhance
the positive vacuum energy density of our closed system causing the system
to behave, thermodynamically, as being controlled by the vacuum energy. So,
one can confidently assume that the original massless particles that existed
during the Casimir regime are basically those which where borne out of
vacuum through the mechanism of the Casimir effect plus the finite
temperature enhancement deduced by Plunien etal. Indeed a similar behavior
to the case of dynamical Casimir effect inside a resonantly vibrating cavity
presented by Plunien etal is obresved here where the number of particles
increases all the time. This interpretation, i.e, the finite-temperature
enhansement of the Casimir energy explain, physically, the behavior of
quantum fields at finite temperatures during the Casimir regime.

\section{References}

\thinspace \ \ [\thinspace 1] N. D. Birrell, and P. C. W.Davies, ''Quantum
Fields in Curved Space'',(Cambridge University Press: Cambridge, U.K 1982)

\ [\thinspace 2] A. A. Penzias, and R. W.Wilson, Astrophys. J 142, (1965)
419.

\ [3] D.T Wilkinsoytler, J. M. O`Meara, and D.Lubin, Physica Scripta T85,
(2000) 12.

\ [4] L. H. Ford, Phys. Rev. D 11, (1975) 3370.

\ [5] J. S. Dowker, and R. Critchley, J. Phys. A 9, (1976) 535.

\ [6] J. S. Dowker, and M. B. Altaie, Phys. Rev. D 17, (1978) 417 .

\ [7] G. Gibbon, and M. J. Perry, Proc. R. Soc. London, A 358, (1978) 467 .

\ [8] T. S. Bunch, and P. C. W.Davies, Proc. R. Soc. London, A 356, (1977),
569 , Proc. R. Soc. London, A 357, (1977), 381, and Proc. R. Soc. London, A 
{\bf 360}, (1978), 117.

\ [9] S. Fulling, Phys. Rev. D 7 (1973) 2850.

[10]. G. Kennedy, J. Phys. A11, (1978), L77.

[11]. B. L. Hu, , Phys. lett. B{\bf 103}, (1981), 331.

[12]. J. S. Dowker, and R. Critchley, Phys. Rev. D 15, (1977), 1484.

[13]. M. B. Altaie, and J. S. Dowker, Phys. Rev. D 18, (1978), 3557.

[14]. G. Plunien, , R. Schutzhold, and G. Soff, , Phys. Rev. Lett. 84,{\bf \ 
}(2000), 1882.

[15]. M. B. Altaie, J. Phys. A 13 (1978) 1603.

[16]. S. Singh and R. K. Pathria, J. Phys. A 17 (1984) 2983.

[17]. M. B. Altaie, and E. Malkawi, J. Phys. A 33 (2000) 7093.

[18]. L. Parker and Zhang, Phys. Rev. D 44 (1991) 2421.

[19]. L. Parker and Zhang, Phys. Rev. D 47 (1993) 2483.

[20]. S. Smith and J. D. Toms, Phys. Rev. D 53 (1996) 5771.

[21] A. Dolgov and Ya. Zeldovich, Rev. Mod. Phys. 53 (1981) 1.

[22] L. Parker, Phys. Rev. 83 (1969) 1057.

\end{document}